\documentclass[aps, prl, reprint, amsmath, superscriptaddress, showpacs]{revtex4-1}
\usepackage{graphicx}
\usepackage{hyperref}

\newcommand{\ve}[1]{\boldsymbol{#1}}

\begin{document}


\title{Emergent collective phenomena in a mixture of hard shapes through active rotation}

\author{Nguyen H.P.\ Nguyen}
\affiliation{Department of Mechanical Engineering, University of Michigan, Ann Arbor, Michigan 48109, USA}
\author{Daphne Klotsa}
\affiliation{Department of Chemical Engineering, University of Michigan, Ann Arbor, Michigan 48109, USA}
\author{Michael Engel}
\affiliation{Department of Chemical Engineering, University of Michigan, Ann Arbor, Michigan 48109, USA}
\author{Sharon C.\ Glotzer}
\email{sglotzer@umich.edu}
\affiliation{Department of Chemical Engineering, University of Michigan, Ann Arbor, Michigan 48109, USA}
\affiliation{Department of Materials Science and Engineering, University of Michigan, Ann Arbor, Michigan 48109, USA}

\date{\today}

\begin{abstract}
We investigate collective phenomena with rotationally driven spinners of concave shape.
Each spinner experiences a constant internal torque in either a clockwise or counterclockwise direction.
Although the spinners are modeled as hard, otherwise non-interacting rigid bodies, we find that their active motion induces an effective interaction that favors rotation in the same direction.
With increasing density and activity, phase separation occurs via spinodal decomposition, as well as self-organization into rotating crystals.
We observe the emergence of cooperative, super-diffusive motion along interfaces, which can transport inactive test particles.
Our results demonstrate novel phase behavior of actively rotated particles that is not possible with linear propulsion or in non-driven, equilibrium systems of identical hard particles.
\end{abstract}


\maketitle


\textit{Introduction.}---Active matter is a rapidly growing branch of non-equilibrium soft matter physics with relevance to fields such as biology, energy, and complex systems~\cite{Marchetti}.
In active matter, dissipative, steady-state structures far-from-equilibrium can emerge in systems of particles by converting energy to particle motility~\cite{Vicsek2012, Marchetti}.
Recent works have reported novel collective behavior not possible with passive matter, such as giant number fluctuations~\cite{Search2003}, clustering~\cite{peruani2006nonequilibrium}, swarming~\cite{dorsogna06, Nguyen2012}, fluid-solid phase separation of repulsive disks~\cite{Bialke2012, fily2012athermal, Redner2013}, and collective rotors~\cite{Wensink2013}. Effective interactions emerging between hard, self-propelled particles were shown to cause phase separation~\cite{tailleur2008statistical,cates2012motility,Redner2013,Stenhammar2013} and coexistence~\cite{Redner2013} in simulations. 
Experimentally, some of these phenomena were demonstrated by driving the system via vibration~\cite{Narayan06072007, kudrolli2008swarming, Roeller2011}, chemical reaction~\cite{palacci2010sedimentation, theurkauff2012dynamic}, and light activated propulsion~\cite{Palacci2013}. To date, most studies have focused on self-propulsion where the constant input of energy to each particle goes directly into translational motion and hence active forces couple to particle velocities.
Converting the input of energy into \textit{rotational} motion, however, does not directly influence translational motility, and couples only to the particles' angular momentum.
We denote such a coupling of active driving forces to angular velocity as \emph{active rotation}.

\begin{figure}[b!]
\centering
\includegraphics[width=\columnwidth]{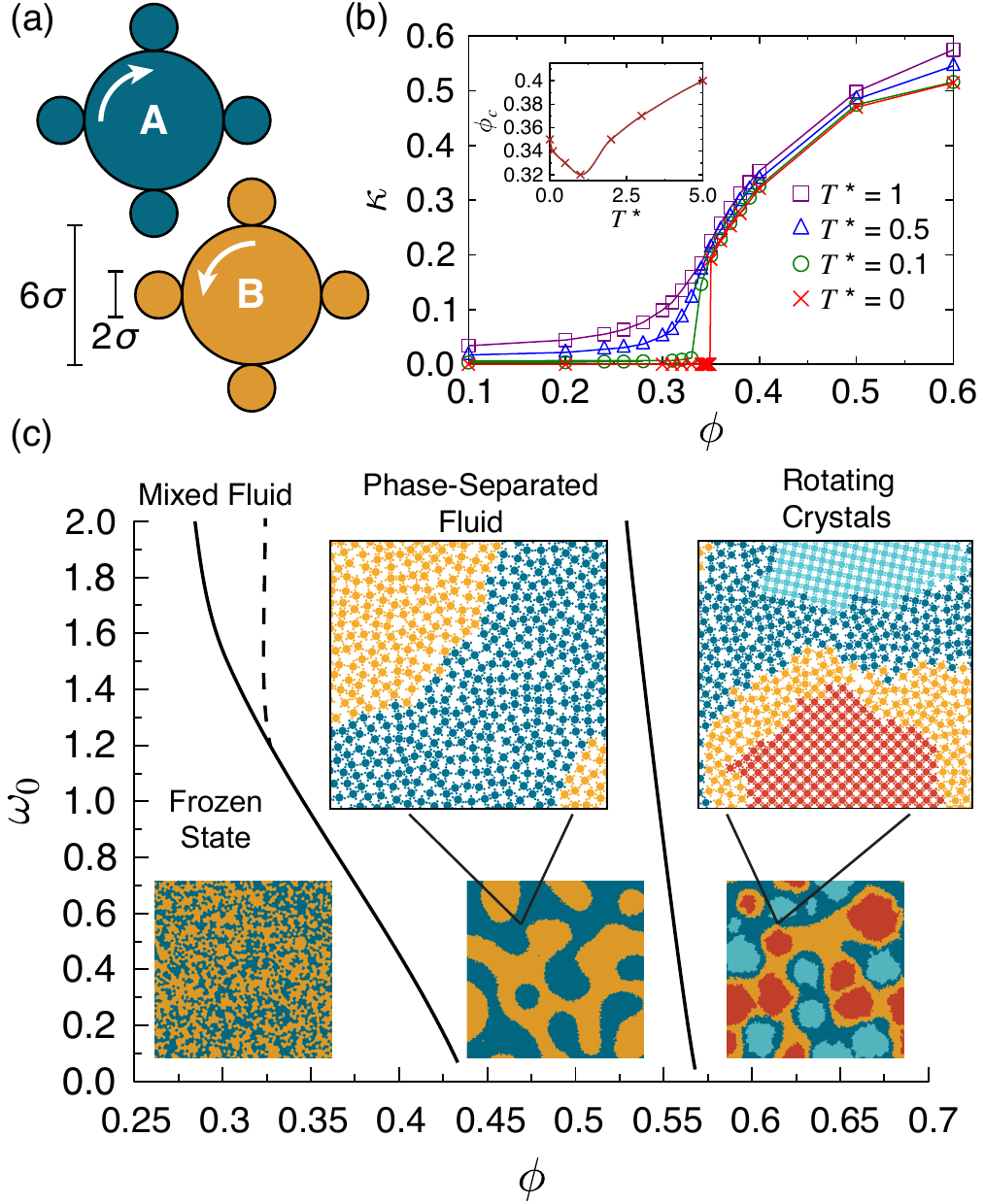}
\caption{
(a)~Schematic diagram of $A$ (clockwise) and $B$ (counterclockwise) spinners.
(b)~The ratio of translational to total kinetic energy $\kappa$ at $\omega_0=1$ indicates the presence of a phase transition.
The inset shows the critical density $\phi_c$ as a function of noise $T^*$. Error bars are smaller than the symbols. 
(c)~Phase diagram based on simulation data (symbols).
Lines are guides to the eye.
Insets show representative snapshots as the system approaches steady state.
A  50:50 mixture of $A$ and $B$ spinners is used in (b,c).
}
\label{panel1}
\end{figure}

Active rotation may be achieved by various methods, \textit{e.g.\ }external magnetic fields~\cite{Grzybowski2002,Yan2012} and optical tweezers~\cite{Grier1997,Moffitt2008,Arzola2012}.
Biological organisms can spin naturally, such as the dancing algae of Ref.~\cite{Drescher2009}.
Actively rotated, macroscopic particles submersed in a fluid~\cite{grzybowski00,Grzybowski2002}, exhibit hydrodynamic interactions~\cite{Ishikawa2006,Drescher2009} that can be attractive or repulsive through the formation of vortices~\cite{Grzybowski2002}.
Actively rotated particles can also exhibit  cooperative self-propulsion~\cite{Fily2012}, synchronization, and self-proliferating spiral waves~\cite{Uchida2010}.
Beyond these examples, however, the potential use of active rotation for pattern formation and self-organization in driven systems has not been investigated systematically.
In this Letter, we show with computer simulations that effective interactions between spinners rotating in the same directions, and between oppositely rotating spinners, emerge due to the active motion itself.
The result is phase separation, rotating crystals, cooperative and heterogeneous dynamics leading to superdiffusive motion, and complex phase behavior.


\textit{Model and methods.}---Our spinner particle is modeled by four peripheral disks of radius $\sigma$ rigidly attached to a central disk of radius $3\sigma$ at each of the compass points, Fig.~\ref{panel1}(a).
The system is governed by a set of coupled Langevin equations for translation and rotation,
\begin{eqnarray}\label{langevin2}
m\frac{\partial \ve{v}_i}{\partial t} &=& \ve{F}_i -\gamma_{t}\ve{v}_i +\ve{F}_i^R,\\
I\frac{\partial \omega_i}{\partial t} &=& \tau_i^{D} + \tau_i - \gamma_{r}\omega_i  + \tau_i^{R},
\end{eqnarray}
where $m$, $I=64m\sigma^2$, $\ve{v}_i$, and $\omega_i$ have the usual meanings of mass, moment of inertia, and translational and angular velocity.
Each spinner is driven by an external torque $\tau_i^D=\pm\tau^D$ of constant magnitude, but with positive sign for clockwise spinners (`$A$') and negative sign for counterclockwise spinners (`$B$').
Spinners are hard particles that interact via a purely repulsive contact potential, resulting in internal forces $\ve{F}_i$ and torques $\tau_i$.
Energy is dissipated through translational and rotational drag coefficients $\gamma_{t}$ and $\gamma_{r}$.
Noise is included via random forces $\ve{F}_i^R=\sqrt{2\gamma_{t} kT}R(t)$ and torques $\tau_i^{R}=\sqrt{2\gamma_r kT}R(t)$ that model a heat bath at temperature $T$.
Random forces are applied directly to the centroid of each spinner.
$R(t)$ are normalized zero-mean white-noise Gaussian processes, which assures thermodynamic equilibrium in the absence of the externally applied torques.

Langevin Dynamics simulations are performed on graphic processing units (GPUs) with our HOOMD-blue software package~\cite{[{}][{. HOOMD-blue web page: \url{http://codeblue.umich.edu/hoomd-blue}}]Anderson2008a} using up to 16,384~spinners (81,920~disks), half of which are driven to spin always clockwise, and the rest driven to spin counterclockwise.
The hard contact is modeled via a Weeks-Chandler-Andersen potential with parameter $\epsilon$~\cite{Weeks1971} shifted to the surface of the spinner such that its range is a small fraction of the particle diameter, thereby approximating ``perfectly'' hard shapes. 
In the low Reynolds number limit, the translational and rotational drag coefficients are related through the Stokes-Einstein and Stokes-Einstein-Debye relations.
If we approximate the spinner as a disk of effective radius $\tilde\sigma$, the relations for disks give $\gamma_r = \frac{4}{3}\tilde\sigma^2 \gamma_t = 100\sigma^2\gamma_t$.
We choose $\sigma$, and $\epsilon$ as the basic units of length and energy, respectively.
The unit of time is $t_0=\sqrt{m\sigma^2/\epsilon}$.
Thermal noise is specified by $T^*=kT/\epsilon$.  Throughout the paper, we report results only for $\tau^D=\gamma_t=1$ unless stated otherwise.



\textit{Phase behavior of spinners.}---Following the random initialization of particle positions and orientations, and equilibration with active rotation turned off, the time evolution of the active system is characterized by an approach towards steady state.
Energy input to rotational motion by the applied torque is transferred to the translational degrees of freedom and then dissipated by drag forces.
In steady state, we observe that the energy balance $\kappa=E_\text{trans}/E_\text{total}$ between translational and total kinetic energy converges.
While the non-driven 2D system has $\kappa=2/3$ as dictated by the equipartition theorem, a value $\kappa<2/3$ quantifies the non-equilibrium character of the system. 
We analyze the behavior of $\kappa$ as a function of density $\phi$ and noise $T^*$.
As shown in Fig.~\ref{panel1}(b), at low density the driven rotational motion dominates translational motion and $\kappa\rightarrow 0$.
With increasing density, the number of collisions increases and $\kappa$ approaches the equipartition value.
Interestingly, we find a phase transition for densities in the range $0.25<\phi_c<0.48$.
For zero noise, the increase in $\kappa$ is sharp, and possibly discontinuous, but becomes less sharp as the noise increases.
Based on the observation of a rapidly increasing length scale in the pair distribution function $g_\text{AB}(r)$ of opposite spinners (discussed below), we identify this transition as the phase separation of the system into $A$-rich and $B$-rich domains. 
As shown in the inset of Fig.~\ref{panel1}(b), increasing the noise from zero at the same value of applied torque initially lowers the critical density $\phi_c$, because particle collisions that facilitate the onset of phase separation are more frequent. 
If the system is too noisy, phase separation is hindered and the trend is reversed: the system phase-separates at a higher density as the noise increases.

For the remainder of the paper, we follow Refs.~\cite{Drescher2011, Wensink2013} and neglect the role of noise by setting $T^*=0$.
In this limit, the dynamics of the system is fully characterized by two parameters, the density $\phi$ and the low-density steady-state angular velocity $\omega_0 = \tau^D/\gamma_r$, which is a measure of activity.
Fig.~\ref{panel1}(c) shows the $\phi$-$\omega_0$ phase diagram at (or near) steady state.  
Movies of the spinner dynamics and the phase separation process can be found in the Supplementary Materials.

At low densities, we find a frozen (absorbing) state~\cite{Hinrichsen2000};
the spinners become stationary and rotate at angular velocity $\omega_0$.
Translational motion does not couple to rotation, the system is non-ergodic, and drag forces dominate.
The few collisions that occur from the initial kinetic energy (random initialization) die out quickly due to dissipation.
Note that our frozen state is different from those observed under oscillatory shear~\cite{Corte2008} or with self-propulsion~\cite{Schaller2011}, where particles retrace their trajectories. 
 
At higher density, the frequency of collisions increases.
When the time interval between collisions is sufficiently short, \textit{i.e.\ }comparable to the characteristic time $\gamma_t/m$ for energy dissipation, non-stop chained collisions can sustain the transfer from rotational to translational energy.
Depending on $\omega_0$, we observe a transition to either a mixed liquid or a phase separated liquid.
While spinners with high activity, $\omega_0\ge1.2$, display two transitions (first to the mixed liquid, then phase separation), these transitions merge for lower $\omega_0$.

As the density increases further, the spinners organize into crystals that rotate collectively about their centers of mass;
particles no longer rotate about their individual centers.
The critical density decreases as the activity increases in agreement with~\cite{Bialke2012,Redner2013}, but not with~\cite{fily2012athermal}.
If crystallization occurs before phase separation is completed, the crystal size is limited by the phase separating domains.
The angular velocity of a crystal decreases with radius, similar to rotors self-assembled from polymers by self-propelled bacteria~\cite{Schwarz-Linek2012a}.
Dynamically, this phase represents a new kind of active crystal  --a rotating crystal-- distinct from the two previously reported types, traveling and resting crystals~\cite{Menzel2013}.

\begin{figure}
\includegraphics[width=\columnwidth]{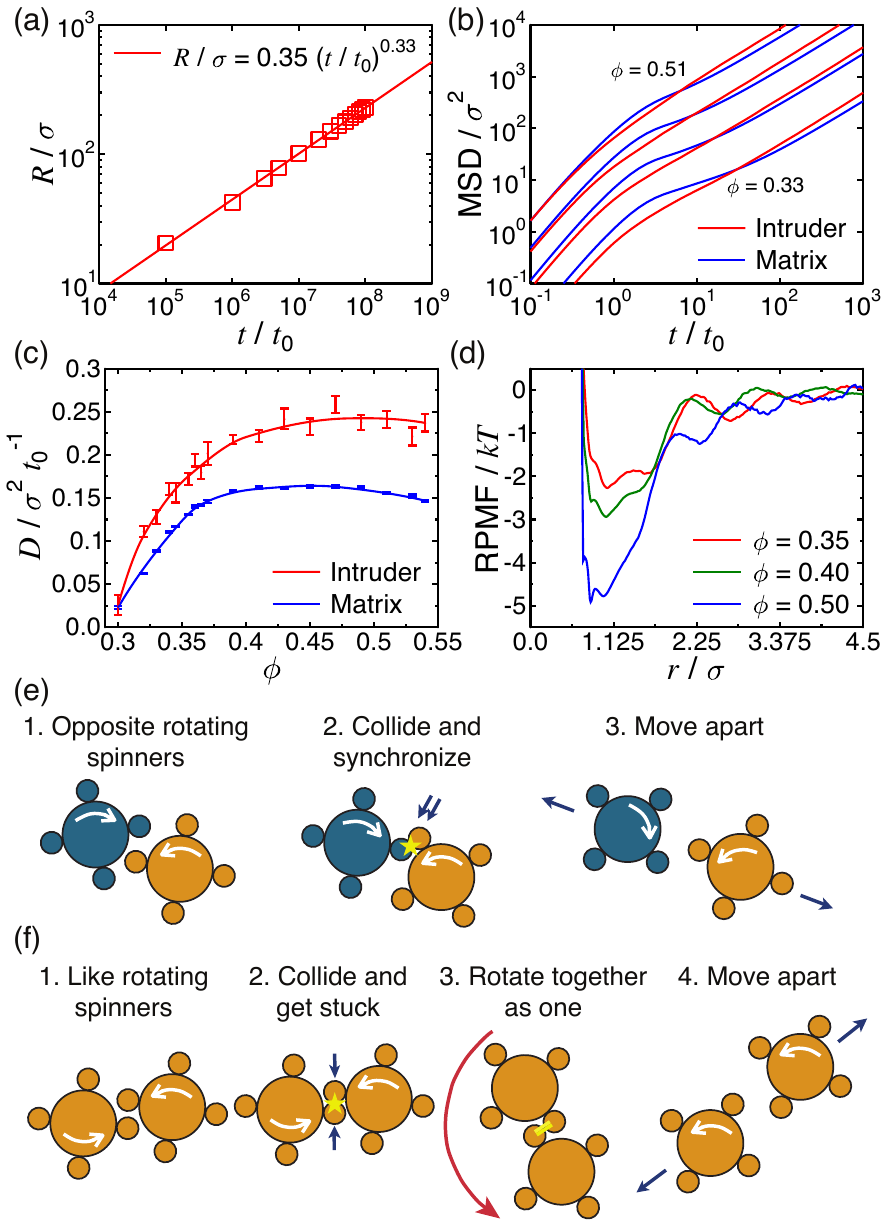}
\caption{
(a)~Domain size growth for a 50:50 mixture of spinners during phase separation ($\phi=0.5$).
(b,c)~Comparison of (b)~mean-squared displacements and (c)~translational diffusion coefficient $D$ for an intruder in a matrix of opposite spinners.
Curves in~(b) at different densities are offset for clarity.
(d)~Relative potential of mean force (RPMF) obtained for $N_A=100$  and $N_B=2$.
In (a-d), activity is set to $\omega_0=1$.
(e,f)~Typical interaction of (e)~two like and (f)~two opposite spinners.}
\label{analysis}
\end{figure}

\textit{Effective interaction between spinners.}---We measure the characteristic domain size as the first zero of $g_\text{AB}(r)$ in Fig.~\ref{analysis}(a).
Domains coarsen over time with an exponent of 1/3, typical for spinodal decomposition in any dimension of a binary mixture in the absence of hydrodynamics~\cite{Jose1983, Revi1993, Glotzer1995}.
The exponent is identical to that measured in a biological system of self-organizing mussels~\cite{Liu2013}, but different from that found in other systems of self-propelled colloids~\cite{Redner2013, Stenhammar2013}.

The origin of the phase separation is investigated with a system of one $B$ spinner (the intruder) in a dense matrix of $A$ spinners.
We compare their mean-squared displacements $\text{MSD}=\langle|\ve{x}_i(t)-\ve{x}_i(0)|^2\rangle$ in Fig.~\ref{analysis}(b).
While matrix particles have higher kinetic energy as seen in the ballistic regime $t<t_0$, the curves cross and diffusion of the intruder is faster for $t\gg t_0$.
The MSD of the matrix has a plateau indicative of caging. 
We extract the translational diffusion coefficient $D$ and plot it as a function of density in Fig.~\ref{analysis}(c).
With increasing density, diffusion speeds up and the gap between intruder and matrix spinners widens.  This differs from a common finding for self-propelled particles that the diffusion coefficient of particles in the dense phase of the phase-separated states decreases to zero as the density increases~\cite{fily2012athermal,cates2012motility,Redner2013,Stenhammar2013}. 
This difference in behavior demonstrates that, although we find that self-rotating and self-propelling systems share some similarities, self-rotating particles exhibit distinctly new phenomena.
Thus the theory developed for self-propelled particles such as in~\cite{tailleur2008statistical,cates2012motility,Stenhammar2013} may not be applicable to self-rotating particles and highlights a need for a more comprehensive theory. 

\begin{figure*}
\includegraphics[width=\textwidth]{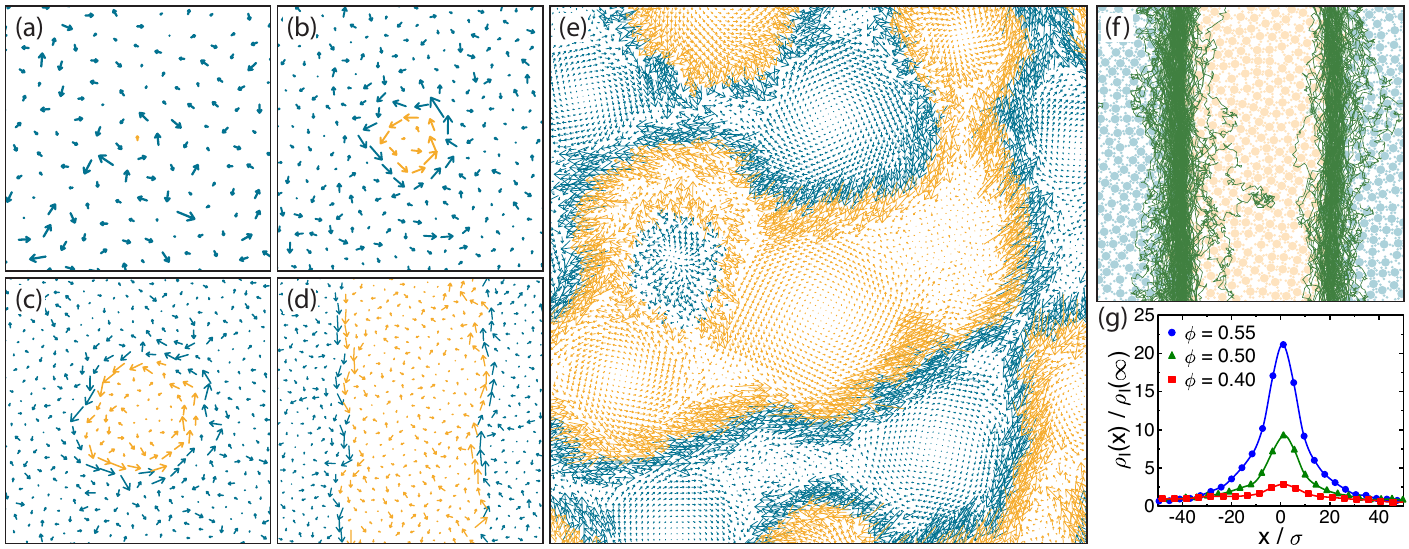}
\caption{
(a)-(d)~Vector plots show the short-time diffusion $\Delta\ve{x}(t)$ in systems of $N=576$ spinners with different numbers of intruders: (a)~one, (b)~$2\%$, (c)~$10\%$ and (d)~$50\%$.
The snapshots (a)-(c) are zoomed in.
(e)~Large system during phase separation at a density where crystallization occurs.
Observation time windows are (a-d)~$10 t_0$ and (e)~$t=100 t_0$.
(f)~Trajectories of inactive particles (green) and (g)~their probability density close to an interface.
In the figure, we use $\omega_0=1$.
Densities are: (a-d,f)~$\phi=0.5$ and (e)~$\phi=0.6$.
}
\label{arrows}
\end{figure*}

We quantify the effective pairwise interaction between $N_B$ intruders among $N_A$ matrix spinners by computing the relative potential of mean force (RPMF) in the limit of a vanishing density of intruders $n_B=N_B/(N_A+N_B)$,
\begin{equation}
V_\text{RPMF}(r) = -kT \lim_{n_B\rightarrow 0} \log \left(\frac{(N_A-1)\rho_\text{BB}(r)}{(N_B-1)\rho_\text{BA}(r)} \right).
\end{equation}
Here, $\rho_{BA}$ and $\rho_{BB}$ are local time-averaged densities for finding a $B$ spinner next to an $A$ and a $B$ spinner, respectively, and the definition guarantees $V(r\rightarrow\infty)=0$.
The RPMF is the average work needed to swap an intruder at infinite separation with a matrix particle at distance $r$.
We find an attractive well in the RPMF of several $kT$ that deepens with increasing density, Fig.~\ref{analysis}(d).
At the critical density $\phi_c=0.35$, the well depth is about $2kT$.
We note that this value of interaction strength is comparable to the attraction strength required to phase separate binary Lennard-Jones liquids~\cite{vlot1997,lamm2001} and to the critical reduced interaction in the 2D Ising model, $2.269 kT$~\cite{onsager1944}.

The microscopic origin of the effective interaction can be understood by comparing pairs of neighboring spinners.
Consider the interaction of two opposite spinners, Fig.~\ref{analysis}(e). The ``teeth" of the gear-like spinners move together for part of the cycle, synchronizing their rotation when in contact due to steric restriction, and then move apart.
Now consider the interaction between two spinners rotating in the same direction, Fig.~\ref{analysis}(f).
Since the tangential velocities at contact are in opposite directions, the spinners momentarily ``stick" sterically.
They cannot spin about their individual axes and instead transfer their angular momentum momentarily to the pair and rotate together for part of the cycle before moving apart.
The consequence is a longer contact time for like spinners compared to opposite spinners, breaking the symmetry between otherwise identical particles and resulting in an emergent, effective attraction (repulsion) between like (unlike) neighboring spinners. 


\textit{Collective dynamics at interfaces and transport.}---As spinners phase separate, they form interfaces separating regions of opposite rotation.
The short-time diffusion $\Delta\ve{x}(t)=\ve{x}(t)-\ve{x}(0)$ is visualized for a density where phase separation, Fig.~\ref{arrows}(a-d), and also crystallization, Fig.~\ref{arrows}(e), occur.
We observe that the translational dynamics of the spinners is heterogeneous and cooperative, in particular at the interfaces.
While diffusion is Brownian in the bulk, spinners speed up significantly by moving linearly along the interface in a super-diffusive manner with $\text{MSD}\propto t^2$. 
As phase separation progresses, the total length of the interfaces decreases, reducing the number of super-diffusive spinners.
Interestingly, we find that both the translational and rotational kinetic energies are uniform across the demixed fluid (see Supplementary Materials).
This means spinners at interfaces do not move faster, but instead move farther in a given time window.
Such dynamical behavior is reminiscent of the string-like dynamical heterogeneity in supercooled liquids and dense colloids~\cite{Kob1997,Donati1998}.

To investigate the possibility of extracting useful work from the active motion of the spinners, we add inactive test particles to the system.
An inactive particle has the same shape, size, and hard interaction as a spinner, but is not subject to a rotational driving torque.
From their trajectories, Fig.~\ref{arrows}(f), we observe that inactive particles diffuse to the interface and get dragged along the 
interface by the current of the active spinners. This observation is confirmed by the density of inactive particles as a function of the distance $x$ to the interface, $\rho_I(x)$, relative to their density in the bulk, $\rho_I(\infty)$, which is strongly peaked at the interface, Fig.~\ref{arrows}(g).
The preference of inactive particles to sit at the moving interface increases with density and could be utilized for collective transport at mesoscopic scales.


\textit{Discussion.}---The phase separation of rotationally driven active particles is realizable in experiment provided the particles are permanently assigned a rotation direction while  being free to move translationally and collide with one another. Applying torques to photo-sensitive spinners through optical trapping is one promising route~\cite{Moffitt2008,Arzola2012} and may even be possible on the colloidal scale where the dynamics can be observed through the microscope~\cite{Grier1997,Palacci2013}. 
In many situations, however, a restriction of the rotation direction is not possible. This is in particular the case for three-dimensional systems.
Our observation of an emergent preferential interaction between like rotating particles suggests a possible alignment~\cite{Uchida2010} of the rotation axes in three dimensions.
Whether self-rotated particles in 3D can synchronize spontaneously into a \emph{nematic spinner phase} (alignment of the rotation axes) remains to be seen.
We further note that we also observe phase separation in preliminary studies with Brownian Dynamics simulations, where the inertial term is absent and the dynamics is dominated by viscous drag forces. This observation raises interesting questions about the presence or absence of local conservation of angular momentum. Finally, we note that the tendency towards phase separation and synchronization will be enhanced by hydrodynamic interactions~\cite{grzybowski00,Ishikawa2006,Drescher2009}. We therefore expect the phenomena reported in this work to occur at lower density and activity levels in experiment that include hydrodynamic effects.


\acknowledgments
This work was supported by the Non-Equilibrium Energy Research Center (NERC), an Energy Frontier Research Center funded by the U.S. Department of Energy, Office of Science, Office of Basic Energy Sciences under Award Number DE-SC0000989.
D.K. acknowledges funding by the FP7 Marie Curie Actions of the European Commission, Grant Agreement PIOF-GA-2011-302490 Actsa.
N.N. also acknowledges the Vietnam Education Foundation for prior support.
We thank P.\ Chaikin, R.\ Larson, R.\ Newman, A.\ Osorio and M. Spellings for useful discussions.

\bibliography{library.bib}

\end{document}